\documentclass{emulateapj}

\slugcomment{Received 2018 March 16; revised 2018 March 30; accepted 2018 March 30}

\shorttitle{A recollimation shock in the $\gamma$-ray NLS1 1H 0323+342}

\shortauthors{Doi, et al.}

\begin{document}

\title{A Recollimation Shock in a Stationary Jet Feature with Limb-brightening in the Gamma-ray Emitting Narrow-line Seyfert~1 Galaxy 1H~0323+342}

\author{Akihiro Doi\altaffilmark{1,2}, Kazuhiro Hada\altaffilmark{3,4}, Motoki Kino\altaffilmark{5,6}, Kiyoaki Wajima\altaffilmark{7}, and Satomi Nakahara\altaffilmark{2,1}}
\altaffiltext{1}{The Institute of Space and Astronautical Science, Japan Aerospace Exploration Agency, 3-1-1 Yoshinodai, Chuou-ku, Sagamihara, Kanagawa 252-5210, Japan}\email{akihiro.doi@vsop.isas.jaxa.jp}
\altaffiltext{2}{Department of Space and Astronautical Science, The Graduate University for Advanced Studies, 3-1-1 Yoshinodai, Chuou-ku, Sagamihara, Kanagawa 252-5210, Japan}
\altaffiltext{3}{Mizusawa VLBI Observatory, National Astronomical Observatory of Japan, Osawa, Mitaka, Tokyo 181-8588, Japan}
\altaffiltext{4}{Department of Astronomical Science, The Graduate University for Advanced Studies (SOKENDAI), 2-21-1 Osawa, Mitaka, Tokyo 181-8588, Japan}
\altaffiltext{5}{Kogakuin University, Academic Support Center, 2665-1 Nakano, Hachioji, Tokyo 192-0015, Japan}
\altaffiltext{6}{National Astronomical Observatory of Japan, 2-21-1 Osawa, Mitaka, Tokyo 181-8588, Japan}
\altaffiltext{7}{Korea Astronomy and Space Science Institute (KASI), 776 Daedeokdae-ro, Yuseong-gu, Daejeon 34055, Republic of Korea}

\begin{abstract}  
We report the discovery of a local convergence of a jet cross section in the quasi-stationary jet feature in the  $\gamma$-ray-emitting narrow-line Seyfert~1 galaxy (NLS1) 1H~0323+342.   The convergence site is located at $\sim7$~mas (corresponding to the order of 100~pc in deprojection) from the central engine.  We also found limb-brightened jet structures at both the upstream and downstream of the convergence site.  We propose that the quasi-stationary feature showing the jet convergence and limb-brightening occurs as a consequence of recollimation shock in the relativistic jets.  The quasi-stationary feature is one of the possible  $\gamma$-ray-emitting sites in this NLS1, in analogy with the HST-1 complex in the M87 jet.  Monitoring observations have revealed that superluminal components passed through the convergence site and the peak intensity of the quasi-stationary feature, which showed apparent coincidences with the timing of observed $\gamma$-ray activities.    
\end{abstract}

\keywords{galaxies: active --- galaxies: Seyfert --- galaxies: jets --- radio continuum: galaxies --- galaxies: individual (1H 0323+342) --- gamma rays: galaxies}


\section{INTRODUCTION}\label{section:introduction} 
The detection of GeV $\gamma$-ray emission from narrow-line Seyfert~1~(NLS1) galaxies by the Large Area Telescope on board the {\it Fermi Gamma-ray Space Telescope} satellite \citep[{\it Fermi} LAT;][]{Abdo:2009} raises issues of the $\gamma$-ray production processes and acceleration mechanisms of relativistic jets in this subclass of active galactic nuclei~(AGNs). 
Radio observations have provided strong evidence of a pole-on viewed relativistic jet by finding very high brightness and rapid variability on the flat-spectrum core associated with the one-side jet \citep[e.g,][]{Doi:2006,Doi:2011a,DAmmando:2013}, which are apparently quite similar to that of blazars \citep[e.g.,][]{Foschini:2015}.

\object{1H 0323+342} is the nearest (redshift of $0.0629$; \citealt{Zhou:2007}) NLS1 among known  $\gamma$-ray-emitting radio-loud NLS1s.  One of the detected $\gamma$-ray flares showed a flux-doubling time of 3~hr, which was the fastest $\gamma$-ray variability ever observed from NLS1s \citep{Paliya:2014,Paliya:2015}.   
Prominent radio flares with a timescale of $\sim 30$~days have been evident only at high frequencies,  
while the source activity is very moderate at lower frequencies \citep[$\la 10$~GHz;][]{Angelakis:2015}.  The $\gamma$-ray/radio correlation for \object{1H 0323+342} has not been investigated in detail.    
One-sided morphology and superluminal components were identified at parsec scales by monitoring observations using very-long-baseline interferometry \citep[VLBI;][]{Fuhrmann:2016,Lister:2016}.  
On the other hand, two-sided radio morphology is seen at kiloparsec scales \citep{Anton:2008}.  These characteristics can be understood analogously to the unified scheme of radio-loud AGNs that considers radio galaxies as non-beamed parent populations of blazars \citep{Doi:2012}.  

The presence of a quasi-stationary feature located at $\sim7$~mas ($\sim120$~pc
\footnote{If we adopt a black hole mass of $M_{\rm BH} \sim 2 \times 10^7 M_\sun$ \citep[][and references therein]{Landt:2017} and an inclination angle of $\sim3$\degr \citep{Abdo:2009} for \object{1H 0323+342}, 1~mas corresponds to 1.2~pc in projection or $4.5 \times 10^5 R_{\rm S}$ in deprojection, where $R_{\rm S}$ is the Schwarzschild radius.  We assume a $\Lambda$CDM cosmology with $H_0=70.5$~km~s$^{-1}$~Mpc$^{-1}$, $\Omega_\mathrm{M}=0.27$, and $\Omega_\mathrm{\Lambda}=0.73$.}
) from the \object{1H 0323+342} nucleus was suggested by \citet{Wajima:2014}.  
Such a structure is reminiscent of the \object{M87} ($z=0.004283$) jets including the HST-1 complex \citep{Biretta:1999}; the upstream end of the feature has been stable at 860~mas (a deprojected distance of 120~pc) from the nucleus \citep{Cheung:2007}.    
The HST-1 complex has been identified as one of the possible very high-energy~(TeV) $\gamma$-ray sites in \object{M87} \citep{Cheung:2007}.    
The location of this $\gamma$-ray activity is much farther from the central engine than previously thought in relativistic jet sources.  
This phenomenon can be explained as a consequence of inverse-Compton upscattering off ambient starlight photons \citep{Stawarz:2006}.   
\citet{Asada:2012} discovered that the cross section of the jet shows locally smaller at the HST-1 in the jet-width profile.  
They discussed the origin of the HST-1 as a consequence of a recollimation shock \citep{Stawarz:2006} due to overcollimation by a pressure imbalance between the jet and ambient medium \citep[e.g.,][]{Gomez:1997,Mizuno:2015} around the sphere of the Bondi radius.  
The plausible $\gamma$-ray-emitting sites on a standing shock located far from the central engine have also been proposed for distant blazars (BL~Lac, \citealt{Marscher:2008}; PKS~1510$-$089, \citealt{Marscher:2010}; OJ~287, \citealt{Agudo:2011}).  

Because of its proximity, the quasi-stationary feature in \object{1H 0323+342} potentially becomes another case in addition to the HST-1 complex in \object{M87}, giving us precious opportunities to make detailed studies for the possible site of recollimation and $\gamma$-ray productions by direct imaging using VLBI angular resolutions.  
In this Letter, we present the observation result that the quasi-stationary feature in the \object{1H 0323+342} has a converging/diverging structure with limb-brightening, which is likely the result of a recollimation/reflected shock.  
Furthermore, from VLBI monitoring data, we show an indication of a possible $\gamma$-ray site on the quasi-stationary feature, which much farther from the \object{1H 0323+342} central engine.

\begin{figure}
\epsscale{1.15}
\plotone{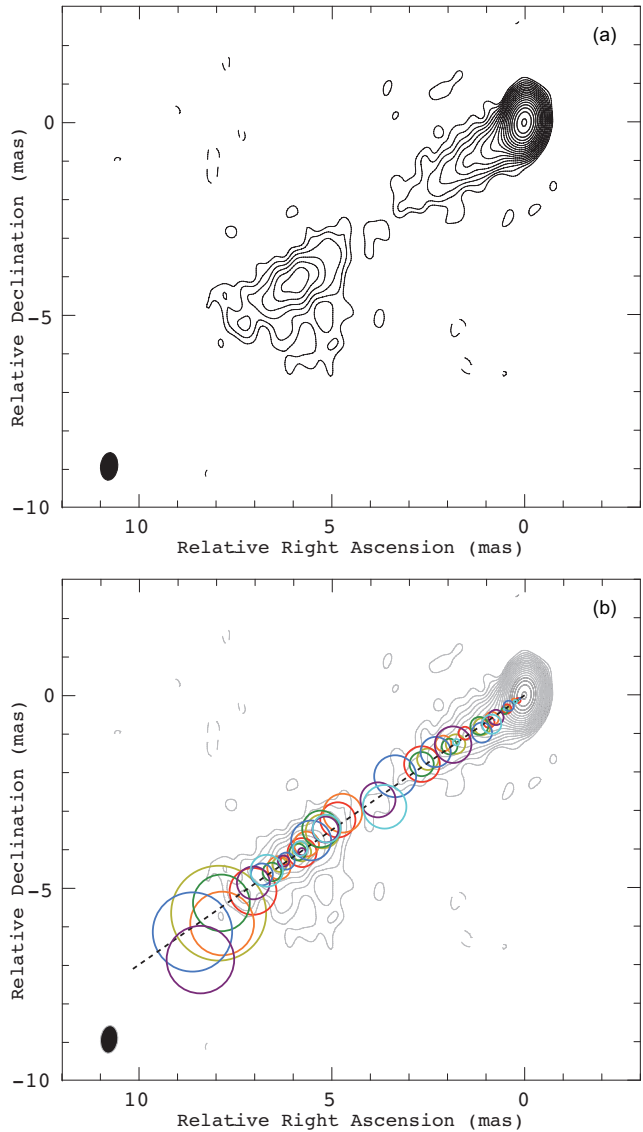}
\figcaption{
Stacked image and result of model fitting for 1H~0323+342 jet.  (a)~The total intensity contour map that was made by stacking images over all eight epochs, provided by the MOJAVE project.  (b)~Overlay plots for the positions and deconvolved sizes of components in the sky with colors depending on epochs.  Black dashed line represents the jet axis at $PA=125\fdg0$.
\label{figure:stacked_image}
}
\end{figure}

\begin{figure}
\epsscale{1.15}
\plotone{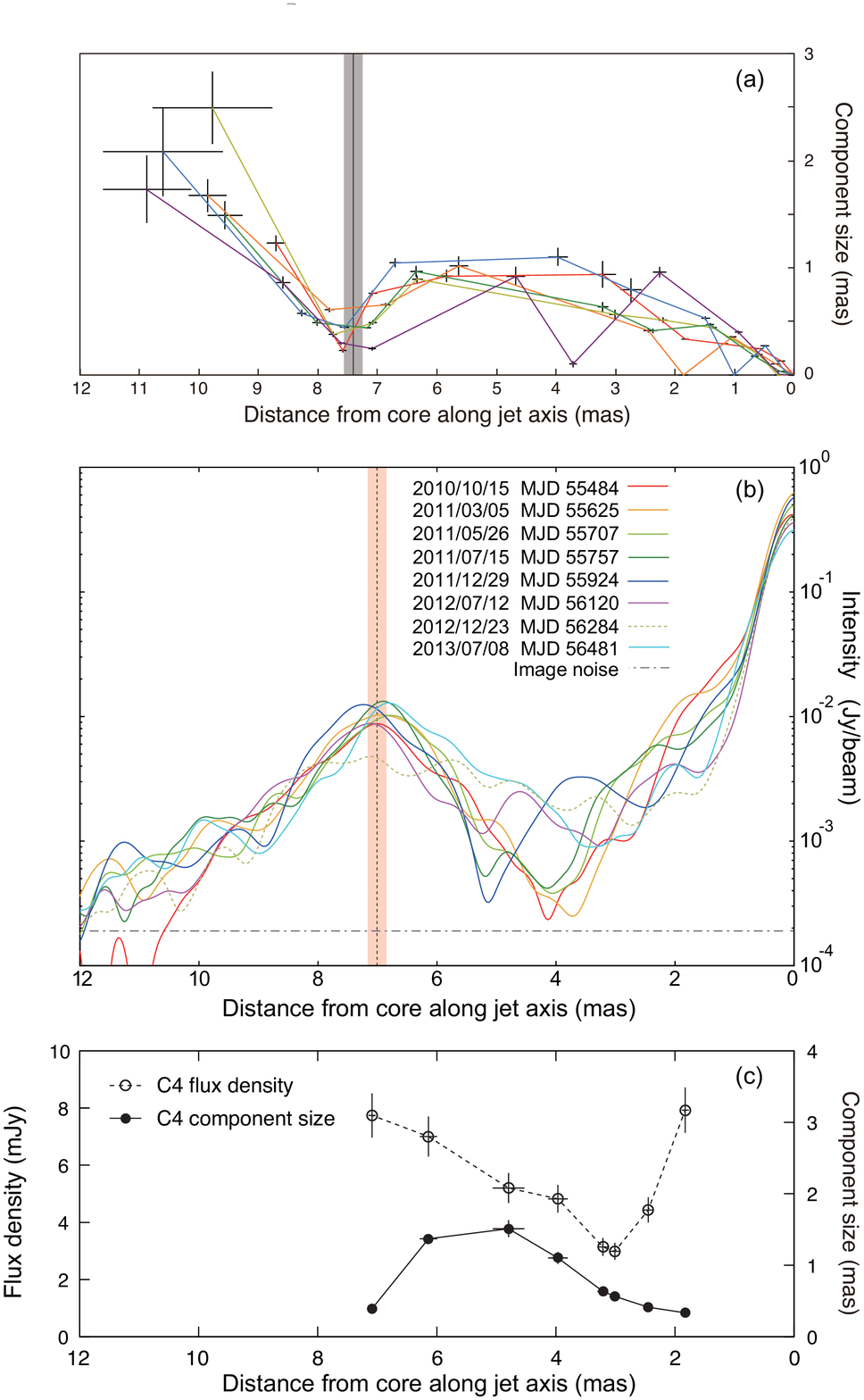}
\figcaption{
Spatial-domain plots for 1H~0323+342 jet components.  (a)~Deconvolved sizes of components with distance from the core.  The vertical solid line and gray shaded belt represent the average and standard deviation of the location of convergence site in the jet, respectively.  (b) Slice profiles of intensity along the jet axis.  Color variations denoted in the panel are shared with the panel Figure~\ref{figure:stacked_image}~(b).  The vertical dashed line and red shaded belt represent the average and standard deviation of the location of intensity peak on the quasi-stationary feature ``S'' in the jet, respectively.  (c)~Evolution of component C4 regarding flux density and size in radius.
\label{figure:spatial-domain}
}
\end{figure}

\section{Data and Data analyses}\label{section:data}
We retrieved all of the calibrated visibility data for \object{1H 0323+342} observed at 15.4~GHz using the very long baseline array~(VLBA) from the Monitoring Of Jets in AGNs with VLBA Experiments (MOJAVE) project \citep{Lister:2005} online page\footnote{\url{https://www.physics.purdue.edu/MOJAVE/sourcepages/0321+340.shtml}}.  
The data were acquired from 210 October~15 to 2013 July~08 (eight epochs in total) in a series of observations coded BL149 and BL178.  
We analyzed the visibilities using {\tt difmap} software \citep{Shepherd:1994}.  Radio structures were established using circular Gaussian model components in {\tt modelfit} procedure.  

\citet{Fuhrmann:2016} reported their results obtained in the same manner; most of the identified components were consistent with our results.  We were not confident of model fitting for the data on 2012 December~23 (the seventh epoch) because of significantly poor data quality compared to the other epochs.  For the present study, we adopt only features cataloged with robust cross-identifications across epochs in the table of the MOJAVE paper \citep{Homan:2015} for the seventh epoch.  
The formal errors of the model fit parameters were estimated to be $\sigma_z = \theta/(2 \times {\rm SNR})$ and $\sigma_r = \theta/{\rm SNR}$ for the position and deconvolved component size, respectively \citep{Fomalont:1999}.   
In the case of very small components and high flux density, this tends to underestimate the error.  We included an additional minimum error of $\theta/10$ to the position error by root-sum-square \citep[e.g.,][]{Doi:2006,Britzen:2010,Lu:2012}.  

We created CLEAN images from the calibrated visibility data.  A standard CLEAN procedure without self-calibration was performed using {\tt difmap}.  These images are used to investigate the intensity profiles of the jet.

\begin{figure}
\epsscale{1.15}
\plotone{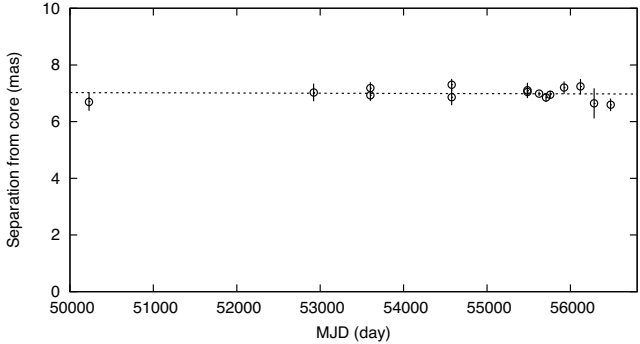}
\figcaption{Historical location of the quasi-stationary feature S.  The data before and after modified Julian date (MJD)~55484~(2010 October~15) come from \citet{Wajima:2014} and the present study, respectively.  The solid line represents fitted proper motion.    \label{figure:historicallocation}}
\end{figure}

\section{RESULTS}\label{section:results}

\subsection{A converging Jet Structure}\label{section:result:convergent}
Pc-scale jet structures were modeled using a series of discrete components.  
Figure~\ref{figure:stacked_image}~(a) shows a stacked image made from images from all the eight epochs.  Figure~\ref{figure:stacked_image}~(b) shows the jet components regarding the positions and deconvolved sizes, overlaid with the stacked image.  There is an emission gap in the jet at a radial distance of $z\sim3$--$5$~mas from the radio core; the intensity subsequently rises in the downstream at $z\sim7$~mas.        
This panel indicates that the evolution of component sizes experiences a local minimum near the local maximum of intensity in the jet.  

Figure~\ref{figure:spatial-domain}~(a) illustrates the component size profiles along jet axis.  The jet initially expands in a parabolic shape after launched from the core.  The jet becomes narrowed at a downstream.  The local minimum is located at $z = 7.40 \pm 0.17$~mas from the core ($d = 0.39 \pm 0.13$~mas in size), which is the average and standard deviation of measurements for one or two local minimums at each epoch.  We call this location the ``convergence site.''   The jet appears to expand again at the downstream of the convergence site.  

The jet-width profile over a wide distance range for \object{1H 0323+342} is separately reported by \citet{Hada:2018}, which also confirms a locally narrowed jet based on multi-frequency (1.4--43~GHz).

\subsection{A Quasi-stationary Feature in the Jet}\label{section:result:quasi-stationary}

Figure~\ref{figure:spatial-domain}(b) exhibits slice profiles of intensity along the jet axis of $PA=125\fdg0$, on the basis of images regenerated with a common restored beam of 0.75~mas, which is the geometric mean of the major and minor axes of original beam sizes.  We found that the intensity maximum in the jet is placed at $z = 6.97 \pm 0.17$~mas from the core, which is the average and standard deviation of the peak positions at the all epochs.  
The location of the convergence site is slightly shifted from that of the intensity peak in the jet feature.

It has been pointed out that this jet feature is stationary for a corresponding component; \citet{Wajima:2014} reported a proper motion of $\mu_\mathrm{app} = -0.115 \pm 0.083$~mas~yr$^{-1}$ during $\sim15$~year.    Figure~\ref{figure:historicallocation} shows the historical locations, including the measurements in our analyses for an additional further eight epochs in MOJAVE data.  We derived a proper motion of $\mu_\mathrm{app} = 0.00 \pm 0.01$~mas~yr$^{-1}$ and a historical position of $7.0 \pm 0.1$~mas for $\sim18$~year.  We found some degree of wobble in distance position during the MOJAVE period.  Hence, we call this feature ``quasi-stationary feature S.''     
  
We recognize the existence of another potential stationary component very close to the core, $z\sim0.3$~mas, as previously pointed out by \citet{Wajima:2014} and \citet{Fuhrmann:2016}.

\subsection{Time Evolution of jet components}\label{section:result:evolution}
\begin{figure}
\epsscale{1.2}
\plotone{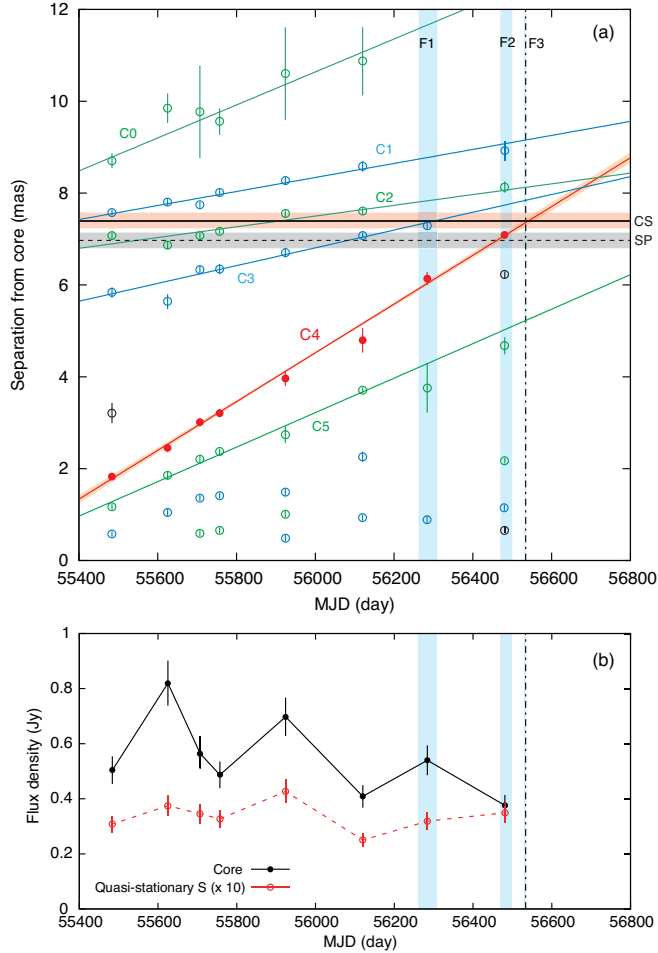}
\figcaption{Time-domain plots for 1H~0323+342.  (a)~Measurements of separations from the core for jet components.  The solid lines represent fitted proper motions for components (the yellow shaded belt on C4 shows fitting error).  The horizontal solid line and red shaded belt represent the average and standard deviation of the location of convergence site (CS) in the jet.  The horizontal dashed line and gray shaded belt represent the average and standard deviation of the location of quasi-stationary feature's intensity peak (SP) in the jet.   (b)~Radio light curves of the core region and quasi-stationary features S in MOJAVE images. 
The periods shaded by light blue and denoted as F1, F2, and F3 represent known $\gamma$-ray activity periods, which were defined by \citet{Paliya:2014}.  The vertical doted-dashed line represents the time of a fast $\gamma$-ray flare at $\mathrm{MJD}=56,534.13$ \citep{Paliya:2014,Paliya:2015}.  
\label{figure:timedomain}}
\end{figure}

Figure~\ref{figure:timedomain}~(b) is the plot for the positions of components in distance from the core.  
The identifications of a series of components and proper motions are similar to those in previous studies using the same MOJAVE data \citep{Fuhrmann:2016,Lister:2016}.  The component C4 has been evidently identified as a superluminal component with the highest apparent speed in this source.  A model with a constant apparent speed fits the measurement points quite well, with $\chi^2/{\rm ndf} = 5.112/6$ and a probability of $p =0.53$, where ${\rm ndf}$ is the number of degrees of freedom.

Figure~\ref{figure:spatial-domain}~(c) shows the evolutions of C4 regarding its flux density and angular size.  This component initially expanded as it became dark, then turned to flare up as it approached the convergence site along the intensity profile of the quasi-stationary feature S.  The other components also showed similar evolution (not shown).  This behavior indicates how the part of the jet is maintained as a quasi-stationary feature.

\subsection{Limb-brightening Structure}\label{section:limb-brightening}
\begin{figure*}
\epsscale{1.15}
\epsscale{1.0}
\plotone{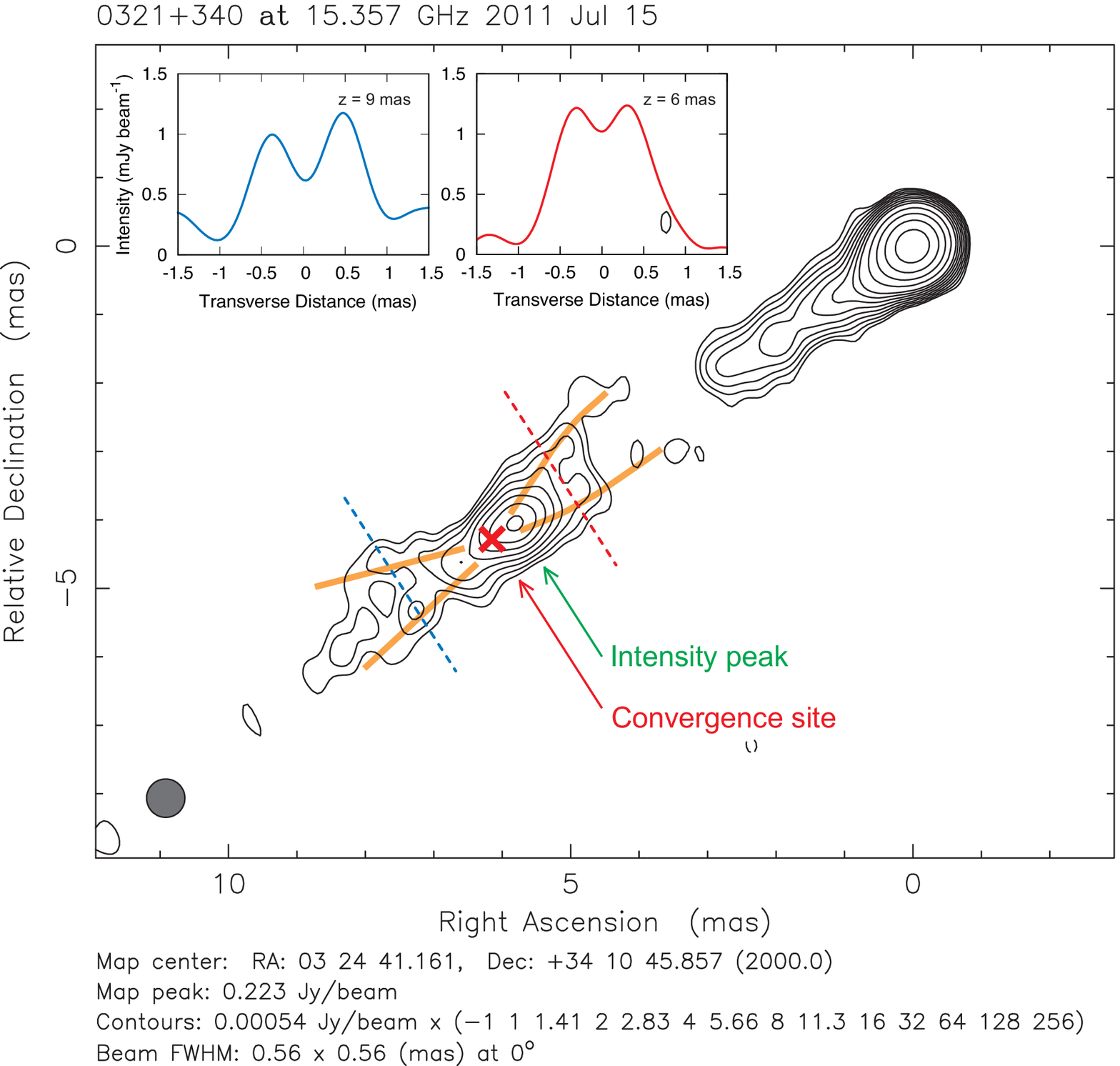}
\figcaption{Jet structure of 1H~0323+342.  Contours represent the total intensity map of MOJAVE data on the epoch $\mathrm{MJD} = 55757$.  Sub-figures show the sliced intensity profiles in transverse direction at the locations $z=6$ (red color) and $z=9$ (blue color) from the core.  Bold lines illustrate a putative sheath layer in the jet as a shocked plasma after a recollimation shock.        
\label{figure:limb-slice}}
\end{figure*}

Figure~\ref{figure:limb-slice} shows the CLEAN image on the epoch of 2011 July~15, which was convolved with a circular restoring beam with 0.56~mas that was equivalent to the minor-axis size of the original synthesized beam.       
We found limb-brightening intensity profiles at both the upstream and downstream with respect to the intensity peak and the convergence site in the quasi-stationary feature S.  We performed a pixel-based analysis by slicing in transverse directions of the jet at $z=6$ and $z=9$ on the image; a double-peaked slice profile is evident in the both cases.     
Similar structures were also apparent at upstream at four epochs and/or downstream at six epochs among the eight epochs.

\section{DISCUSSION}\label{section:discussion}

Our analyses of the radio imaging data for \object{1H 0323+342} have provided us with two key results. 
The first key result is the discovery of a recollimation site as the converging shape in the pc-scale jet, by measurements for the deconvolved sizes of jet blobs.  Furthermore, the intensity peak of the quasi-stationary jet feature spatially coincides with, or slightly upstream of, the convergence site.  
The second key result is the finding of limb-brightening structures at both the upstream and downstream with respect to the intensity peak and the convergence site in the quasi-stationary feature.  
These critical sites are considerably distant from this NLS1 central engine.  If we adopt a black hole mass of $M_{\rm BH} \sim 2 \times 10^7 M_\sun$ \citep[][and references therein]{Landt:2017} and an inclination angle of $\sim3$\degr~\citep{Abdo:2009}, the projected distance $z\sim7$~mas to the quasi-stationary feature S from the nucleus corresponds to $\sim 120$~pc or $\sim 6 \times 10^7 R_{\rm S}$ in deprojection.

\begin{table}
 \caption{Estimated MJDs of Passing Through the Two Critical Sites.}
 \begin{center}
   \begin{tabular}{lccc}
    \hline\hline
Component & Quasi-stationary S   & Convergence Site   \\
 & (MJD)   & (MJD)   \\
 (1)   & (2) & (3)  \\
\hline
C3    & $    56144    \pm    145    $ & $    56417    \pm    155    $ & \\
C4    & $    56479    \pm    45    $ & $    56561    \pm    48    $ & \\
C5    & $    56998    \pm    101    $ & $    57112    \pm    109    $ & \\
    \hline
   \end{tabular}
  \end{center}
\tablecomments{Column.~(1) ID of components; Column~(2) estimated MJD of passing through the location of the intensity peak of the quasi-stationary feature S; Column~(3) estimated MJD of passing through the location of the convergence site.\label{table:passingMJD}}
\end{table}

\subsection{Analogy with M87}\label{section:analogywithM87}
These characteristic jet structures in \object{1H 0323+342} are reminiscent of the \object{M87} jets in pc scales and the HST-1 complex at a deprojected distance of 120~pc.  The inner radio jet of \object{M87} showing a limb-brightened structure \citep[e.g.,][]{Kovalev:2007a}.  \citet{Asada:2012} discovered a jet structure locally convergent at the location of HST-1, which is the transitional boundary of a jet acceleration region with a parabolically expanding streamline and a deceleration region with a conical expansion \citep[e.g.,][]{Nakamura:2013,Asada:2014}.  \object{1H 0323+342} also shows a parabolic shape in the inner jet and a hyperbolic/conical shape at larger scales \citep{Hada:2018}.  
The location of the upstream end in the HST-1 complex in \object{M87} has been stable to within $\sim 2$~mas at 860~mas from the core \citep{Cheung:2007}.  Although there has been no report on the limb-brightened structure near the HST-1 complex, superluminal ejections from the HST-1 showed a variation in the position angle \citep{Giroletti:2012}, which may be because the downstream of the HST-1 has a wide opening angle.  It is proposed that HST-1 is the result of a recollimation shock, a stationary shocked region resulting from the overcollimation in the jet, which explains the broadband flaring activity including TeV $\gamma$-ray emission \citep{Stawarz:2006,Cheung:2007}.

\subsection{Limb-brightened Structure with a Recollimation Shock}\label{section:limbbrightening}

We propose that the combination of the observed limb-brightened structure and the convergence site in the quasi-stationary feature in the \object{1H 0323+342} jet can be explained in the framework of the recollimation shock.  We consider that the limb-brightening at the upstream of the convergence site is possibly a radiation layer of shocked materials flowing between a contact discontinuity that separates from the external medium and the recollimation shock front.  The low-brightness inner layer is considered to be an unshocked spine flow.  
Such a picture has been demonstrated by \citet{Bromberg:2009} and \citet{Bodo:2018} to explain the activity in a small region located at a considerable distance from the central engine in several blazars and \object{M87}.  The shape of the post-shocked dissipation layer after the recollimation shock is very similar to the limb-brightening at the upstream of the convergence site in \object{1H 0323+342}.  \citet{Bromberg:2009} and \citet{Bodo:2018} also revealed that significant pressure loss caused by radiative cooling can lead to the jet being focused on a tiny cross-sectional radius by external pressure.  This picture can explain well the fact that the intensity peak and convergence site are adjacent to one another, as observed in \object{1H 0323+342}.  The limb-brightening at the downstream in the \object{1H 0323+342} jet can be caused by emission from the combination of a shocked outer layer with a relatively high $\Gamma$ and inner layer strongly decelerated through a reflected shock, as shown in \citet{Bodo:2018}.

\subsection{Possible Coincidences of $\gamma$-ray Activity and Jet-passing Events through the Critical Sites}\label{section:gamma-concidence} 

We also discuss possible correlations between $\gamma$-ray activities and passing events of the superluminal component through the intensity peak of the quasi-stationary feature S and the convergence site.  This is the possible third finding in the present study (Figure~\ref{figure:timedomain} (b)).  
The most striking event is that the component C4 reached the intensity peak of S during the known GeV $\gamma$-ray active phase F2 ($\mathrm{MJD} = 56470$--$56500$; \citealt{Paliya:2014}).   Furthermore, the subsequent $\gamma$-ray active phase F3 ($\mathrm{MJD} = 56531$--$56535$) coincides with the component C4's arrival at the convergence site.  The estimated dates of passing through the two critical sites are listed in Table~\ref{table:passingMJD}.    

Table~\ref{table:passingMJD} also presents corresponding dates for the components C3 and C5.  The date of C3 crossing the convergence site was coincident with the previous $\gamma$-ray active phase F1 ($\mathrm{MJD} = 56262$--$56310$).  The proper motion of C5 is expected to cross the two critical sites at $\mathrm{MJD} \sim 56900$--$57200$.  At this period the {\it Fermi} LAT daily and weekly quick-look light curves\footnote{\url{https://fermi.gsfc.nasa.gov/ssc/data/access/lat/msl\_lc/}} (not shown) show recurrence in $\gamma$-ray in the duration of $\mathrm{MJD} \sim 57000$--$57300$.  
As far as these data are concerned, there are apparent associations between $\gamma$-ray emission and passage events of jet components.  
The recollimation site in the quasi-stationary feature is a possible candidate for observed $\gamma$-ray emission in \object{1H 0323+342}.

X-ray spectral modeling suggested a dominance by nonthermal jetted (inverse-Compton) emission during both F2 and F3, whereas in the quiescent (and F1) phase there is a significant dominance by thermal corona \citep{Paliya:2014}.  Given the hour-scale light curve in F3 and its estimated $\gamma$-ray luminosity, a good fraction of the total kinetic energy of the jet was converted into radiation \citep{Paliya:2014}.  
They proposed that the dissipation sites for all of these $\gamma$-ray phases are located $\sim 100 R_{\rm S}$ ($\sim 0.002$~pc) from the nucleus, with region sizes of $\sim0.0005$~pc based on the one-zone broadband spectral modeling.   
On the other hand, the convergence site has a cross-sectional diameter of $\sim0.4$~mas ($\sim0.3$~pc).  Some compact substructures confined to the narrowing cross section are required to reconcile the $\gamma$-ray productions at the quasi-stationary feature.   

The C4 showed a progressively increasing flux density by a factor of three (Figure~\ref{figure:spatial-domain} (c)).  This indicates how the quasi-stationary feature keeps stationary, i.e., as a result of the aggregation of passing components that brighten as they approach the convergence site.  However, these flux densities were, at most, at the mJy level.  On the other hand, radio flares seen in the F-GAMMA light curves based on single-dish monitoring were on the order of 1~Jy \citep{Angelakis:2015}.   
We examined the $\gamma$-ray/radio correlation on spatially resolved light curves using the MOJAVE images (Figure~\ref{figure:timedomain}~(b), a similar plot was also presented by \citealt{Fuhrmann:2016}).      
However, we found no clear evidence of $\gamma$-ray/radio correlation, either in the core region or the quasi-stationary feature. 
These inconclusive results may be due to the MOJAVE's sparse sampling ($\geq 50$~days) compared to the flare's time scales ($\la30$~days) in the total radio flux \citep{Angelakis:2015}.  Searching for component emergence events for both the core and the stationary feature helps to identify the  $\gamma$-ray-emitting site (Doi, A.\ et~al.\ in preparation).  The superluminal ejections from the stationary location in the HST-1 complex were observed during the 2005 TeV $\gamma$-ray event in M87 \citep{Cheung:2007,Giroletti:2012}.

\acknowledgments
This research has made use of data from the MOJAVE database that is maintained by the MOJAVE team \citep{Lister:2009a}.


\end{document}